
\overfullrule=0pt
\line{}
\hfill {BROWN-HET-884}

\hfill {SUTDP/92/70/9}

\hfill {March 92/Esfand 70}

\vskip 1.0in

\centerline{{\bf TUNNELING IN ANISOTROPIC COSMOLOGICAL
MODELS}\foot{Research supported by Sharif University of Technology}}
\vskip .50in

\centerline {R.Mansouri and M.Mohazzab }

\centerline { Department of Physics, Sharif University of Technology }

\centerline { P.O.Box 11365-9161, Tehran, Iran }

\vskip 2.0in

\centerline { ABSTRACT }

     Tunneling rate is investigated in homogenous and anisotropic cosmologies.
The calculations is done by two methods:
 Euclidean and Hamiltonian approaches. It is
found that the probability decreases exponentialy as anisotropy is increased.

\endpage

\chapter{{\bf Introduction }}

  New COBE-results of the background microwave radiation strongly supports the
inflationary scenarios in the cosmological models [Silk 1992]. Paradoxially,
the inflationary-universe scenarios should teach us how the present state of
the universe, such as the isotropy of the microwave radiation, is independent
of
the initial state of the universe [Guth 1980, Brandenberger 1985, Olive 1990].
Usually the inflationary models are based on isotropic
Friedman-Robertson-Walker(FRW) metrics. Phase transitions and bubble nucleation
are very much discussed in this respect [Coleman 1977,Callan and Coleman and
Coleman and DeLuccia 1980]. Specifically, the rate of bubble nucleation is very
important in the old inflationary scenario.

Wald [1983] , Turner [1986] and Stein-Schabes [1986] , discuss the
role of anisotropy in new inflation and show that anisotropy
will damp out the
inflationary era. Jensen and Ruback [1989] discuss the decay of false
vacuum in Kantowski-Sachs anisotropic model, using the Euclidean
approach. They calculate
numerically the rate of bubble nucleation and come to the conclusion that the
rate of nucleation is slower than the Coleman-DeLuccia isotropic case.

 The idea of old inflation enters also the extended inflationary model[La and
Steinhardt 1989, Holman et al 1990]; but the role of anisotropy is not known in
these models.

  In this paper we consider homogenous but anisotropic Bianchi-I [Ellis and
McCallum 1969]
 models. We
calculate the rate using the Euclidean approach first. But the calculation is
only valid for models having a minimum amount of anisotropy, Besides this, the
Euclidean approach is not very reliable [Goncharov and Linde 1986]. Therefore
we use also the Hamiltonian approach to calculate the rate of bubble
nucleation. The result, which is also valid for the limit of vanishing
anisotropy, indicate that the the anisotropy enhance the difficulties due to
the graceful exit problem of the old scenario.
\vskip 1.0in

\chapter{{\bf Tunneling in homogenous and anisotropic cosmolgies: Euclidean
approach}}

\vskip .50in

 Here we will consider Bianchi-I compact cosmology as an anisotropic model.
The total action including the effect of geometry and matter field is

  $$S= -{1\over {16\pi G}}\int d^4 x \sqrt {-g} R- \int d^4 \sqrt {-g}
\L_{field},\eqno\eq $$

where $\L_{field}$ is the lagrangian density for a scalar matter field

$$L_{field} = -{1\over 2 }g^{\mu \nu }\partial_{\mu} \phi \partial_{\nu} \phi
-V(\phi ), \eqno\eq $$

and $V(\phi )$ is a two well scalar potential shown in fig.1.

The metric for Euclidean Bianchi-I model is

    $$ds^2= d\tau ^2 + X_1 ^2(\tau )dx_1 ^2+ X_2 ^2(\tau )dx_2 ^2+ X_3 ^2(\tau
)dx_3 ^2,  \eqno\eq$$
where $\tau $ is the Euclidean time and we have assumed the manifold to be
compact;
i.e. $x_1 , x_2 , x_3 \in [0,2\pi ]$ so that any space section is a 3-torous.
we assume the tunneling to be homogenous, i.e.
the matter field  is a function of Euclidean time only

      $$\phi \equiv \phi (\tau )            \eqno\eq$$

  It is more suitable to work with the new metric variables $a$, $\beta_- $ and
$\beta_+ $ defined through the following relations

$$X_i =ae^{\beta_i} \ \ \ \ \ \ \ i=1,2,3,          \eqno\eq $$

where    $\sum_{i=1 }^3 \beta_i =0 $, and

       $$\beta_1 =\beta_+ +\sqrt 3 \beta_-, $$
       $$\beta_2 =\beta_+ -\sqrt 3 \beta_-,          \eqno\eq$$
       $$\beta_3 =-2\beta_+, $$

For the above metric the scalar curvature is

$$R= -6[\dot \beta_+^2 +\dot \beta_-^2 +{\ddot a\over a}+ ({\dot a\over a})^2].
\eqno\eq $$

  Now The Euclidean action can be written as

   $$S_E =\int d^4 x a^3\lbrace {1\over 2}{\dot \phi ^2}+ V(\phi )+{3\over
\kappa
}[-({\dot a \over a})^2 +(\dot \beta_+ ^2 + \dot \beta_- ^2)]\rbrace,
\eqno\eq$$

where: $\kappa =8\pi G $.

   The integrand in eq.(2.8) does only depend on $\tau $ and therefore the
space part can be integrated to a constant. As this constant is irrelevant for
our future discussion, we can assume it to be equal to $1 $. It then leads to

  $$S_E =\int d\tau a^3\lbrace {1\over 2}{\dot \phi^2}+ V(\phi )+ {3\over
\kappa }[-({\dot a \over a})^2 + (\dot \beta_+^2+\dot \beta_-^2)]\rbrace.
\eqno\eq $$
$\beta_{\pm} $ are cyclic coordinatees, therefore

  $${\delta S_E \over {\delta \dot \beta_{\pm} }}=C_{\pm},   \eqno\eq
$$
where $C_{\pm} $ are constants and measure the anisotropy of the model.
  Now from (2.9) and (2.10) we have

$$ \dot \beta_{\pm} = ({\kappa \over 6 })a^{-3} C_{\pm}. \eqno\eq$$

 The field equations are

$$ \ddot \phi + 3{\dot a \over a}\dot \phi = V'(\phi ),   \eqno\eq$$

$$({\dot a \over a})^2 - (\dot \beta_+^2 +\dot \beta_-^2 )= {\kappa \over
3}({1\over 2 }\dot \phi^2 -V(\phi )). \eqno\eq $$

 Substituding eq.(2.13) into the action we find that

$$S_E =2\int d\tau a^3 V(\phi ). \eqno\eq$$

Now by using eq.(2.11), (2.13) can be written as

$$({\dot a \over a})^2 - {\kappa \over {6a^6}}C^2 = {\kappa \over 3}({1\over
2}\dot \phi ^2 - V(\phi )), \eqno\eq $$
where $C^2=C_-^2 +C_+^2$ is a measure of the anisotropy of the model.

  Defining $b=a^3$, equation (2.15) can be
written as

 $${1\over 9}({\dot b \over b})^2 -{\kappa \over {6b^2}}C^2 = {\kappa \over
3}({1\over 2}\dot \phi^2 -V(\phi )). \eqno\eq $$

    Evaluating eq.(2.16)
at a minimum of the potential $V(\phi_{min} )$, where $\dot \phi \simeq 0 $
we obtain

$$\dot b^2= {\kappa \over 2}C^2-3\kappa b^2 V(\phi_{min}). \eqno\eq $$

  For the semiclassical calculation to be valid the scale factor $a$ must be

                 $$a\gg l_{plank},  \eqno\eq$$

  Applying eq.(2.18) to (2.17) a lower limit for $C$ can be found:

    $$ C^2\gg 6l_{plank}^{2\over 3}V(\phi_{min}). \eqno\eq $$

  Therefore the calculation of this section is valid only for large
anisotropies
and
it is not possible to go to the limit $C\rightarrow 0 $. Hence, this method
cannot be applied to flat FRW model.

   Now by defining $b= {\kappa C\over 2}r$  the field eqs.(2.12) and (2.16) can
be written in the following forms

 $$\dot r^2 = 1+ 3\kappa r^2 ({1\over 2}\dot \phi ^2 - V(\phi )),
\eqno\eq $$
 $$\ddot \phi +{\dot r\over r}\dot \phi = V'(\phi ).  \eqno\eq $$

Substituding all the redefinition in the action (2.14) its final form can be
written as

 $$S_E ={\kappa \over 2}C\int d\tau r V(\phi ). \eqno\eq $$
$r$ and $\phi $ are the solution of eqs.(2.20) and (2.21) and clearly
independent
of $C$. So eq.(2.22) means that the decay rate $\Gamma $ decreases exponentialy
as anisotropy increases:

$$\Gamma \sim exp-[{\kappa C\over 2} A ]   \eqno\eq $$
where $A$ is the integral appearing in eq.(2.22), and independent of $C$.
\medskip

\chapter{{\bf Hamiltonian approach to bubble nucleation}}
\medskip

   In the pevious section we saw that the flat FRW limit i.e., $C\rightarrow 0
$, cannot be found by Euclidean method.
  There are in general, other shortcomings in the Euclidean approach [Goncharov
and linde 1986,Widraw 1991]

   Here we use the more reliable method of Hamiltonian approach based
on ADM formalism [Goncharov and Linde 1986]. First we write the metric for the
Bianchi-I model in the following form

 $$ds^2 = -N(t)^2 dt^2+ X_1 ^2 dx^2 + X_2 ^2 dy^2 + X_3 ^2 dz^2,
\eqno\eq $$
where we have left the lapse function $N(t)$ for later use.

  Writing the metric as

$$ ds^2=-N(t)^2 dt^2 + a(t)^2 \sum_{i=1}^3 e^{\beta_i} dx_i ^2,
\eqno\eq  $$
where $\beta_i $s are choosen as in eqs.(2.4), we obtain the curvature scalar
 for the metric (3.1) in the form

 $$R= {6\over N^2 }[\dot \beta_- ^2 + \dot \beta_+ ^2 + ({\ddot a\over a}) +
({\dot a\over a})^2- {\dot N \dot a \over {Na}}]. \eqno\eq$$

 Now the action consists of a geometrical part

 $$S_G =\int dt a^3 N {1\over {16\pi G }}R, \eqno\eq $$
and a scalar field contribution

 $$S_f =\int dt a^3 N({1\over {2N^2}}\dot \phi ^2 -V(\phi )),
\eqno\eq $$

  The total Lagrangian, after an integration by part in the
geometrical contribution of the action, can be written as

  $$L_t ={3\over {8\pi G}}{a^3\over N}[\dot \beta_+^2 +\dot \beta_-^2 - ({\dot
a \over a})^2]+ a^3 N({1\over {2N^2}}\dot \phi^2 - V(\phi )].
\eqno\eq $$

 From the Lagrangian (3.6) we obtain for the canonical momentum

 $$\Pi_N = {\partial L_T \over {\partial {\dot N }}}=0,$$
 $$\Pi_{\beta_{\pm} }= {6\over \kappa }{a^3\over N}\dot \beta_{\pm},$$
 $$\Pi_{\phi }= {a^3\over N }\dot \phi,\eqno\eq $$
 $$\Pi_{\Omega }=-{6\over \kappa }{a^3\over N}\Omega, $$

where $\Omega = {\dot a\over a}   $  .

 The Hamiltonian can be written in the form

 $$H=N\lbrace {1\over 2} a^{-3} \Pi_{\phi }^2 + a^3 V(\phi )+ {1\over
12}{\kappa \over a^3}[(\Pi_{\beta_+ }^2 + \pi_{\beta_- }^2)- \Pi_{\Omega }^2
]\rbrace. \eqno\eq$$

 The above system can be quantized by assignments

 $$\Pi_{\phi }\rightarrow {1\over i}{\delta \over {\delta \phi }}, . . .  ,
\eqno\eq $$

 The Schrodinger-wheeler-DeWitt(SWD) equation of the wave function is
obtained by applying the classical constraint

 $${\delta H \over {\delta N }}= 0, \eqno\eq $$
on the wave function $\Psi (\beta_\pm , \Omega , \phi ) $:

$$[-{\partial ^2 \over {\partial \phi ^2}}- {\kappa \over 6}({\partial ^2 \over
{\beta_- ^2}}+ {\partial ^2 \over {\beta_+ ^2}}- {\partial ^2 \over {\partial
\Omega ^2}})+ 2a^6 V(\phi )]\Psi = 0,  \eqno\eq$$

  Now $\Psi $, that is the probability amplitude of a metric configuration, can
be separated as follows

  $$\Psi =\mu (\beta_+ , \beta_- )\eta (\phi , \Omega ).  \eqno\eq $$

 For $\mu $ we obtain after some trivial calculation

 $$\mu (\beta_+ , \beta_- )= e^{iC_+ \beta_+ + iC_- \beta_- },
\eqno\eq $$
where $C_\pm $ are, as before, the measure of anisotropy of the model

 Putting (3.12) in (3.11) and using (3.13) we get the following partial
differential equation for $\eta $

 $$[{\kappa \over 6}{\partial ^2 \over {\partial \Omega^2}}- {\partial^2 \over
{\partial \phi^2}}+ {\kappa \over 6}C^2+ 2a^6 V(\phi )]\eta (\phi , \Omega )=
0  \eqno\eq $$

 Now, the SWD partial differential equation (3.14)
can be solved by the method of Banks Bender and Wu [1973]
(BBW).

  Consider the following two dimentional problem

 $$[-({\partial ^2 \over {\partial x^2}}+ {\partial ^2 \over {\partial y^2}})+
U(x,y)]\Psi (x,y)= 0, \eqno\eq $$

  In $(x-y)-$plane there are many paths connecting any initial state
to a final state. But there is just one path along which the probability
is
maximum. This classical path is called the most probable escape
path (MPEP). We define two parameter $s$ along the MPEP, and $n$ perpendicular
to it. writting $\Psi $ as

   $$\Psi (x,y)= A(x,y)e^{-F(x,y)}, \eqno\eq $$
 $S, V$ and $A$ are expanded with respect to $n$ as follow

 $$ F= F_0(s)+ n^2F_2(s)+... ,\eqno\eq $$
 $$ V= V_0(s)+ nV_1(s)+ n^2V_2(s)+ ... ,\eqno\eq $$
 $$ A= A_0(s)+... .   \eqno\eq $$

  It is easily seen that

  $$F_0'^2= V_0 \eqno\eq $$

Therefore, an approximate solution for (3.15) can be written as

$$\Psi \simeq exp(-\int ds{\sqrt V_0 }), \eqno\eq $$

  Now to solve eq.(3.14) we define the potential $U(\Omega ,\phi )$ as

$$U(\Omega ,\phi )= {\kappa \over 6}C^2+ 2e^{6\Omega }V(\phi ),$$
and tunneling would be from $\phi_{false} $ to $\phi_{true} $.
  Now following the same processes explained
 above, the first term of the expantion of $U (\Omega ,\phi ) $
can be written as

 $$U_0= {\kappa \over 6}C^2+ B, \eqno\eq $$
where $B$ is a function of MPEP parameter $s$ but independent of $C$ .
Therefore, the approximate solution of eq.(3.14) can be written as

  $$\Psi (\Omega , \phi )\simeq exp(-\int ds \sqrt {{\kappa \over 6}C^2 + B
} ). \eqno\eq  $$

  Taking the initial and final states as $\Psi_{false} $ and $\Psi_{true} $,
the decay rate $\Gamma $ can be easily found:

 $$\Gamma = {\vert \Psi_{true} \vert \over {\vert \Psi_{false}\vert }}=
e^{-{\int_{\phi_{false}}^{\phi_{true}} ds\sqrt {{\kappa \over 6}C^2+
B}}}.\eqno\eq $$

  The tunneling probability is therefore exponentialy decreasing as $C$
increases. Note that the result (3.24) is valid also for the limit
$C\rightarrow 0$, i.e. for the flat FRW model. Although, the exact value of
$\Gamma $ depends on the MPEP, but the qualitative behaviour of it is the same
as the result of the Euclidean approach, as far as the decrease of $\Gamma $
with the increase of $C$ is concerned.

\endpage

\chapter{{\bf Conclusions}}

\medskip

 We have investigated the effect of anisotropy on the rate of bubble nucleation
in the Bianchi-I models using the Euclidean and Hamiltonian approachs. The
Hamiltonian approach leads to a result which is also valid as the anisotropy
tends to zero, i.e. for the flat FRW case. Clearly the Euclidean method does
not apply to the FRW case. Our result shows that irrespective of the method of
calculation, the anisotropy decrease the nucleation rate. Hence the graceful
exit problem of the old inflation is enhanced in the case of anisotropic
Bianchi-I cosmology.

   The role of anisotropy in the extended inflationary model is not known yet.
We are investigating the rate of nucleation in anisotropic solutions of
Brans-Dicke cosmology.

\endpage

{\bf Refrences }

\medskip

   Banks T, Bender C, Wu T T 1973 Phys. Rev. D{\bf 8} 3346; {\bf 8}, 3366

   Brandenberger R H Rev. Mod. Phys. 1985 {\bf 57} 1

   Coleman S 1977 Phys. Rev. D {\bf 15 } 2929

   Callan C and Coleman S 1977 Phys. Rev. D {\bf 16} 1762

   Coleman S and DeLuccia F 1980 Phys. Rev. D {\bf 21 } 3305

   Ellis G F R and McCallum M A H 1969 Comm. Math. Phys. {\bf 12} 108

   Goncharov A S and Lide A D 1986 Sov. J. Part. Nucl. {\bf 17 } 369

   Guth A H 1981 Phys. Rev. D {\bf 23} 347

   Jensen L G and Ruback P J 1989 Nuc. Phys. {\bf B325 } 660

   Jensen L G Stein-schabes J A 1986 Pys. Rev. D {\bf 34} 931

   Holman R, Kolb E W, Vadas S L, Wang Y and E Weinberg 1990 Phys. Lett. {\bf
 B237} 37

   La D and Steinhardt P J 1989 Phys. Rev. Lett. {\bf 62 } 376

   Olive K A 1990 Phys. Rep. {\bf 190} 307

   Silk J 1992 Nature {\bf 356} 741

   Turner M S, Widrow L M 1986 Phys. Rev. Lett. {\bf 57} 2237

   Wald R M Phys. Rev. D 1983 {\bf 28} 2118

   Widrow L M 1991 CITA preprint

\end